# Component Outage Estimation based on Support Vector Machine

Rozhin Eskandarpour, Amin Khodaei
Department of Electrical and Computer Engineering
University of Denver
Denver, USA
rozhin.eskandarpour@du.edu, amin.khodaei@du.edu

*Abstract*—Predicting power system component outages in response to an imminent hurricane plays a major role in pre-event planning and post-event recovery of the power system. An exact prediction of components states, however, is a challenging task and cannot be easily performed. In this paper, a Support Vector Machine (SVM) based method is proposed to help estimate the components states in response to anticipated path and intensity of an imminent hurricane. Components states are categorized into three classes of damaged, operational, and uncertain. The damaged components along with the components in uncertain class are then considered in multiple contingency scenarios of a proposed Event-driven Security-Constrained Unit Commitment (E-SCUC), which considers the simultaneous outage of multiple components under an *N-m-u* reliability criterion. Experimental results on the IEEE 118-bus test system show the merits and the effectiveness of the proposed SVM classifier and the E-SCUC model in improving power system resilience in response to extreme events.

## NOMENCLATURE

*Indices and Sets*:
- $b$    Index for buses.
- $B$    Set of components connected to bus $b$.
- $i$    Index for generation units.
- $s$    Index for scenarios.
- $S$    Set of outage scenarios.
- $t$    Index for time.
- $\beta$    Index for training examples in SVM.

*Parameters*:
- $c$    Penalty factor of soft margin SVM.
- $DR_i$    Ramp down rate of unit $i$.
- $DT_i$    Minimum down time of unit $i$.
- $F_i(.)$    Generation cost function of unit $i$.
- $M$    Large positive constant.
- $P_i^{min}$    Minimum generation capacity of unit $i$.
- $P_i^{max}$    Maximum generation capacity of unit $i$.
- $R$    System reserve.
- $UR_i$    Ramp up rate of unit $i$.
- $UT_i$    Minimum up time of unit $i$.
- $x_l$    Line reactance of line $l$.

*Variables*:
- $a_{lb}$    Element of line $l$ and bus $b$ at line-bus incidence matrix.
- $D_{bt}$    Load at bus $b$ at time $t$.
- $g$    Bias of hyperplane from the origin
- $h(x)$    Hyperplane separating training examples (the output of the classifier).
- $I_{it}$    Commitment state of unit $i$ at time $t$.
- $LC_{bts}$    Load curtailment at bus $b$ at time $t$ in scenario $s$.
- $P_{its}$    Real power generation of unit $i$ at time $t$ in scenario $s$.
- $PL_{lts}$    Real power flow of line $l$ at time $t$ in scenario $s$.
- $T_{it}^{on}$    Number of successive ON hours of unit $i$ at time $t$.
- $T_{it}^{off}$    Number of successive OFF hours of unit $i$ at time $t$.
- $UX_{its}$    Outage state of unit $i$ at time $t$ in scenario $s$; 0 if on outage, otherwise 1.
- $UY_{lts}$    Outage state of line $l$ at time $t$ in scenario $s$; 0 if on outage, otherwise 1.
- $v_b$    Value of lost load at bus $b$.
- $w$    Normal vector to the hyperplane
- $x_\beta$    Training example $\beta$.
- $y_\beta$    Class labels of training example $\beta$.
- $\Delta_i$    Permissible power adjustment of unit $i$.
- $\theta_{bts}$    Phase angle of bus $b$ at time t in scenario $s$.

## I. INTRODUCTION

Weather events are the number one cause of power system outages in the United States resulting in millions of dollars damage every year to critical infrastructure, businesses, and emergency networks [1]. Among the different types of extreme weather events, hurricanes are the most frequent in the United States [2]. The power grid resilience in response to these imminent hurricanes can be significantly improved by (a) an efficient forecasting of the likely hurricane impacts on the grid and to have an idea of which components will be damaged, and (b) an optimal schedule of available generations in case of several power component outages. These two factors can potentially improve pre-event planning and post-event recovery practices by preparing the system to address likely component outages and further response to the outages once the hurricane is passed.

A case study on hurricane planning and rebuilding the electrical infrastructure along the Gulf Coast for hurricane Katrina is presented in [3], where the grid resiliency and its interdependency are estimated with data for Hurricane Katrina

This work has been supported in part by the U.S. National Science Foundation under Grant CMMI-1434771.

regarding power delivery and telecommunications, collected post-landfall. Post-disaster restoration is also an important part of the power system resilience studies. In [5] and [6], a stochastic pre-hurricane model is proposed which aims at managing the available resources before the extreme event, as well as a deterministic post-hurricane model to manage the system after the extreme event through a proactive recovery. In [6] an optimal restoration scheduling to minimize the load curtailment in the case of an extreme event is proposed. The proposed model is based on AC power flow constraints and aims at grid restoration after extreme events by using lost load data. In [7] an optimal repair schedule, system configuration, and unit commitment model to restore the damaged grid infrastructure through a post-disaster decision-making model is proposed. A general informative case study of a variety of models and algorithms regarding emergency response logistics in distribution networks is provided in [8]. The study in [9] predicts the outages in response to extreme events and investigates several data-driven measurements, such as the total number of component outages, potential effects on customers, the type of component outages, and their geographic distribution and duration. The study is based on large databases of outages in five hurricanes in Carolina. In [10], to forecast the power outage durations an ensemble learning method for regression is provided which are applied for outages resulted from several hurricanes such as Dennis, Katrina, and Ivan. In [11], a prediction model to estimate hurricane-related power outages is proposed and claimed to be viable along the U.S coastline. This prediction model is applied to publicly available data and used in estimating several historic storms, such as hurricane Sandy and typhoon Haiyan.

The relatively considerable number of hurricanes in the United States have resulted in a substantial amount of data. Machine learning and data mining methods can therefore be viable tools to analyze these data and accordingly enhance power system planning and recovery in response to imminent hurricanes. In this paper, a multi-class Support Vector Machine (SVM) is proposed to classify power system components states after the hurricane into three classes (i.e., operational, outage and uncertain) based on the distance of each component from the center of the hurricane and the category of the hurricane. In the proposed problem three sets of data are available which need to be classified into three groups using SVM, which is a binary classifier. As a result, the classical SVM is extended using one-versus-one multi-classification. During the evaluation to predict the state of each component, each component is evaluated by three classifiers where the majority response of the classifier determines the state of the component. The obtained component outages from the proposed model are further integrated into a developed event-driven security-constrained unit commitment (E-SCUC) model to find the optimal schedule of available resources that minimizes the operation cost while enhancing grid resilience.

Unlike the existing work on outage prediction and extended outage consideration in SCUC, including the previous work of authors in [12] and [13], this paper uses a multi-class SVM to determine uncertain component states in addition to outage and operational. The merit behind considering uncertain states is that in practical applications there is limited information about the impact of the hurricane on all components and therefore considering uncertain states can increase the prediction accuracy and would help with achieving a better scheduling solution. To further consider the uncertain states, the previously defined N-m reliability criterion is extended in this paper to an N-m-u reliability criterion in which u represents the number of uncertain components on outage.

The rest of the paper is organized as follows. Sections II and III respectively present the model outline and the formulation of the proposed outage estimation model and E-SCUC. Section IV presents numerical simulations based on the IEEE 118-bus test system. Section V concludes the paper.

## II. PROPOSED COMPONENT OUTAGE ESTIMATION MODEL

The proposed model has three stages. At first, an SVM model is trained to classify the components into three classes (damaged, operational, and uncertain). The model is trained on multiple features for each element (wind speed, distance to the center of the hurricane). Since historical data of the impact of previous hurricanes on power system components are limited, synthetic data are generated and used to train the model. During scheduling for the hurricane, the category and the path of the potential hurricane that is heading toward the power system are forecasted, which however can be obtained from weather forecasting agencies. According to the forecasted hurricane, the state of each component is determined by the trained SVM model. Once the state of each system component is estimated, the E-SCUC problem is solved to provide an optimal and resilient schedule of the resources in response to the hurricane.

It is common to use the *N*-1 criterion for reliability studies in power systems, where *N* in the total number of components in the system. The *N*-1 criterion simply states that the system needs to adequately and reliably supply loads in case of a single component outage at any given time. Although extremely useful, this criterion cannot be directly applied to the system to ensure resilience, since more than one component in response to a hurricane may be out of service. In [12], an extended criterion, i.e., *N-1-m* is proposed which simultaneously considers the power system security in response to the single component outage, i.e., an *N-1* criterion, and also in response to the outage of *m* components in impacted regions. In order to consider uncertainty in component failure prediction, this paper employs a further extended criterion, i.e., *N-m-u*, where *m* is the number of components that are predicted to be on outage during the extreme event, and *u* is the number of components in uncertain class.

### A. SVM Method for Outage Prediction

The state of a component in response to the hurricane is considered to be either damaged (i.e., on outage), operational (i.e., in service), or uncertain (i.e., there is possibility of outage but it is completely certain). In order to classify the state of each component, different features can be considered. In this paper, the wind speed and the distance of the each component from the center of the hurricane are investigated as two major features. SVM [14] is applied to determine the decision boundary between different classes and accordingly determining the state of each component.

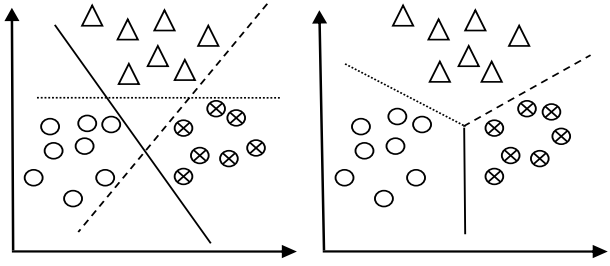

Figure 1. One-versus-all and one-versus-one multiclassification

An SVM can generally classify data into two classes by determining the proper hyperplane that separates training examples of one class from the other one. The best hyperplane is considered as the hyperplane with a clear gap that is as wide as possible. The optimal hyperplane parameters can be found by the following optimization problem:

$$\min \frac{1}{2}\|w\|^2 + c\sum_{\beta=1}^{m} \varepsilon_\beta$$
$$s.t. \quad y_\beta\left(w^T x_\beta + g\right) \geq 1 - \varepsilon_\beta, \quad \beta = 1,......,m \quad (1)$$
$$\varepsilon_\beta \geq 0, \quad \beta = 1,......,m$$

where $w$ is the normal vector to the hyperplane separating training examples, $|g|/\|w\|$ is the perpendicular distance from the hyperplane to the origin, and $c$ is a penalty parameter allowing separating nonlinear examples. This is a quadratic programming problem which can be solved by Lagrange duality. By solving the duality, the final hyperplane only depends on the support vectors (i.e., sample points that are in the margin) and SVM needs to find only the inner products between the test samples and the support vectors.

The idea of the maximum-margin hyperplane is based on the assumption that training data are linearly separable, which is not the case in many practical applications. In order to apply SVM to nonlinear data, kernel methods can be used [14]. The idea of kernel method (or kernel trick) is to map input features to a higher dimension that can be linearly separable and can fit the maximum-margin hyperplane in the transformed feature space. Kernel trick simply states that for all $x_1$ and $x_2$ in the input space X, a certain function $K(x_1, x_2)$ can be replaced by the inner product of $x_1$ and $x_2$ in another space. For example, a polynomial kernel of degree $d$ can be defined as:

$$K(x_\alpha, x_\beta) = (x_\alpha . x_\beta)^d \quad (2)$$

SVMs are inherently two-class classifiers. The traditional way to do multiclass classification with SVMs is to use one-versus-rest classifiers (commonly referred to as one-versus-all) and to choose the class which classifies the test data with the greatest margin. Another strategy is to build a set of one-versus-one classifiers and to choose the class that is selected by the most classifiers (majority voting). In one-versus-one approach, if $g$ is the number of classes, then $g(g-1)/2$ classifiers are constructed, and each one trains data from two classes. Fig. 1 shows the difference between one-versus-all and one-versus-one approaches in a three class problem.

*B. E-SCUC Problem Formulation*

The objective of the E-SCUC problem is defined as:

$$\min \sum_t \sum_i F_i(P_{it0}, I_{it}) + \sum_t \sum_s \sum_b vLC_{bts} \quad (3)$$

where the first term is the operation cost in normal system operation (which includes the generation cost and startup/shutdown costs) and the second term is the cost of unserved energy during contingency scenarios (which is defined as the value of lost load times unserved energy). The value of lost load is defined as the average cost that each type of customer - residential, commercial, or industrial - is willing to pay in order to avoid power supply interruptions [15]. Assuming *UX* and *UY* as outage states for generation units and transmission lines, respectively, the proposed objective function is subject to the following operational constraints:

$$\sum_{i \in B} P_{its} + \sum_{l \in B} PL_{lts} + LC_{bts} = D_{bt} \quad \forall b, \forall s, \forall t \quad (4)$$

$$P_i^{\min} I_{it} UX_{its} \leq P_{its} \leq P_i^{\max} I_{it} UX_{its} \quad \forall i, \forall s, \forall t \quad (5)$$

$$P_{its} - P_{i(t-1)s} \leq UR_i \quad \forall i, \forall s, \forall t \quad (6)$$

$$P_{i(t-1)s} - P_{its} \leq DR_i \quad \forall i, \forall s, \forall t \quad (7)$$

$$T_{it}^{on} \geq UT_i(I_{it} - I_{i(t-1)}) \quad \forall i, \forall t \quad (8)$$

$$T_{it}^{off} \geq DT_i(I_{i(t-1)} - I_{it}) \quad \forall i, \forall t \quad (9)$$

$$\sum_i P_i^{\max} I_{it} \geq D_t + R_t \quad \forall i, \forall t \quad (10)$$

$$0 \leq LC_{bts} \leq D_{bt} \quad \forall i, \forall s, \forall t \quad (11)$$

$$|P_{it0} - P_{its}| \leq \Delta_i \quad \forall i, \forall s, \forall t \quad (12)$$

$$-PL_l^{\max} UY_{lts} \leq PL_{lts} \leq PL_l^{\max} UX_{lts} \quad \forall l, \forall s, \forall t \quad (13)$$

$$\left| PL_{lts} - \frac{\sum_b a_{lb} \theta_{bts}}{x_l} \right| \leq M(1 - UY_{lts}) \quad \forall l, \forall s, \forall t \quad (14)$$

Load balance equation (4) ensures that the total injected power to each bus from generation units and line flows is equal to the total load at that bus. Load curtailment variable ($LC_{bts}$) is further added to the load balance equation to ensure a feasible solution when there is not sufficient generation to supply loads (due to an outage of power system components). Load curtailment will be zero under normal operation conditions. Generation unit output power is limited by its capacity limit and will be set to zero depending on its commitment and outage states (5). Generation units are further subject to prevailing technical constraints including ramp up and down rate limits (6)-(7), and minimum up and down time limits (8)-(9). The system operating reserve requirement is represented in (10). Load curtailment at each bus is constrained by the total load on that bus (11), and the change in a unit generation is further constrained by the maximum permissible limit between normal and contingency scenarios (12). Transmission line capacity limits and power flow constraints are modeled by (13) and (14), respectively, in which the outage state is included to model the line outages in contingency scenarios. Further details can be found in [12], [13].

The E-SCUC problem is formulated using mixed-integer linear programming (MILP). CPLEX 12.6 [16] is used to solve the MILP problem. CPLEX uses a branch and cut algorithm which solves a series of LP sub-problems. In addition, multi-threading can be applied to CPLEX solver which makes the MILP faster and can automatically make adjustments when memory is limited.

## III. NUMERICAL SIMULATIONS

### A. SVM performance

As historical data for the past extreme hurricanes at component level are limited, a set of synthetic data is generated to train the SVM model as 300 samples in outage state, 300 samples in the operational state, and 150 samples in the uncertain area. To define the synthetic data, Saffir-Simpson Hurricane Scale [17] is used to generate wind speed features of the synthetic data. In particular, the generated samples follow a normal distribution function of one-minute sustained wind-speed of different categories with a small Gaussian noise. Initial studies on the generated synthetic data showed that majority of the components in an anticipated hurricane of category 4 & 5 will be on outage, the components in a hurricane of category 1 will be operational, and the majority of the components in a hurricane of categories 2 & 3 will belong to the uncertain area, which verifies that the generated data conform to realistic conditions. The features are normalized to [0, 1] based on the maximum considered values of wind speed and distance. Fig. 2 shows the generated synthetic data.

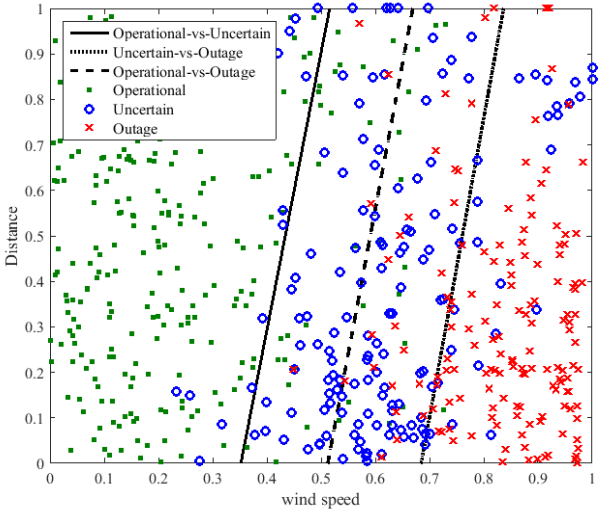

Figure 2. Generated samples for three classes (operational, damaged, and uncertain), and separating hyper-planes of Linear SVM with c=1 in a one-versus-one multi-classification.

A cross-fold validation is a common approach to evaluate a classifier on a set of data. The cross-fold validation splits the data into $q$ subsets, trained on ($q$-1) subsets, and evaluated on the subset that is left in the training. This process is performed $q$ times (such that the classifier is evaluated on all samples). The final classification accuracy is the average of classification accuracies on all folds. In this paper, five folds are defined on the generated set of data ($q$=5). Since our dataset is relatively small, an off-the-shelf SVM model implemented in LibSVM [18] is used in this paper. There are other SVM solvers and techniques such as Liblinear [19] which is optimized for large-scale linear classification and regression.

As the number of attributes (features) in the studied case is smaller than the number of observations (samples), the probability of overfitting to the training data is low. A series of hyper-parameter "c" with various well-known kernels are examined to evaluate its performance and show the sensitivity of the employed SVM to its parameters. In each setting, a cross-fold validation with 5 folds is used to evaluate the performance, and the average accuracy of classification is reported in Table I.

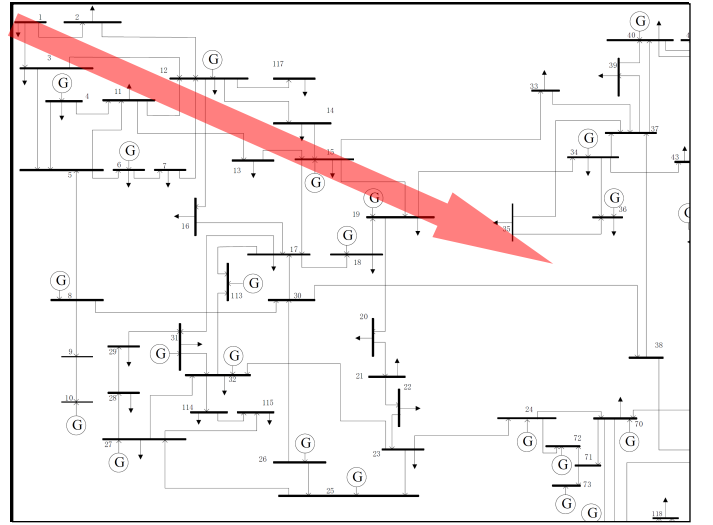

Figure 3. Part of the IEEE 118-bus test system and the forecasted hurricane passing through a hypothetical path

As it is shown, SVM with linear kernel and c=1 offers the best performance among other settings. The average overall classification accuracy of the proposed classification model is 85.3%. The relatively small variance (~2%) in the accuracy of the SVM under various hyper-parameters indicates that the proposed method is not sensitive to its hyper-parameters and is not over-fitted to the training data. Fig. 2 shows the decision boundary of SVM with linear kernel in a one-versus-one multi-classification approach. During the evaluation, to predict the state of each component, each component is evaluated by the three classifiers shown in Fig. 2, and the majority response of the classifier determines the state of the component. Table II shows the confusion matrix of classifying operational and outage components into three states. As it is demonstrated, the proposed method can classify the outage components from the operational components with a high accuracy.

TABLE I. ACCURACY OF SVM WITH VARIOUS HYPER-PARAMETERS "C" AND KERNELS USING 5-FOLD CROSS-VALIDATION

| Kernel | c=0.1 | c=1 | c=10 | c=100 |
|---|---|---|---|---|
| Linear | 84.5 | 85.3 | 85.1 | 85.01 |
| Quadratic | 84.7 | 84.7 | 84.5 | 78.0 |
| Cubic | 84.3 | 82.9 | 78.0 | 71.0 |
| Gaussian | 81.6 | 83.6 | 82.1 | 81.1 |

TABLE II. CONFUSION MATRIX OF CLASSIFYING SYSTEM COMPONENTS DURING EXTREME EVENT (NUMBER OF SAMPLES AND PERCENTAGE)

| | | Predicted | | |
|---|---|---|---|---|
| | | Operational | Uncertain | Outage |
| Actual | Operational | 274 (91.3%) | 25 (8.4%) | 1 (0.3%) |
| | Outage | 1 (0.3%) | 26 (8.7%) | 273 (91.0%) |

### B. E-SCUC with Extended Outage

The proposed E-SCUC problem is applied to the standard IEEE 118-bus test system [20]. A hurricane is assumed to pass through a hypothetical path as shown in Fig. 3. The components in the path and its neighboring areas are classified into three categories of operational, outage, and uncertain, according to the wind speed and the distance to the center of the hurricane. Five components were classified in the outage class (lines 1, 17, 18, 20, and 30) and six components were categorized in the uncertain class (lines 7, 8, 13, 16, 22, and generation unit 1). Table III shows the list of components examined in this case study, the estimated distance from the center of hurricane and the wind speed. Three cases are studied to evaluate the role of considering uncertain area in power system scheduling:

**Case 1**: In this case, only the components in the outage class are considered as possible contingencies and the components in the uncertain area are considered as operational. The operation cost is obtained as $1,041,040. No load curtailment has occurred in this case, so the cost of unserved energy is zero, and the system is secure against this eight component outage.

**Case 2**: In this case, the components in the uncertain class are considered as outages. The operation cost is obtained as $1,085,516. A total of 1.53 MWh load curtailment has occurred in this case.

TABLE III. CLASSIFCATION OF COMPONENTS IN THE STUDIED CASE

| Component | Distance (normalized) | Wind speed (normalized) | Class (prediction) |
|---|---|---|---|
| Line 1 | 0.632 | 0.838 | Outage |
| Line 17 | 0.120 | 0.762 | Outage |
| Line 18 | 0.423 | 0.704 | Outage |
| Line 20 | 0.351 | 0.656 | Outage |
| Line 30 | 0.235 | 0.563 | Outage |
| Line 7 | 0.650 | 0.632 | Uncertain |
| Line 8 | 0.361 | 0.543 | Uncertain |
| Line 13 | 0.532 | 0.562 | Uncertain |
| Line 16 | 0.463 | 0.501 | Uncertain |
| Line 22 | 0.342 | 0.423 | Uncertain |
| Unit 1 | 0.432 | 0.394 | Uncertain |
| Line 60 | 0.640 | 0.437 | Operational |
| Line 48 | 0.497 | 0.395 | Operational |
| Line 47 | 0.401 | 0.254 | Operational |
| Line 44 | 0.302 | 0.202 | Operational |
| Line 57 | 0.252 | 0.184 | Operational |
| Line 45 | 0.740 | 0.156 | Operational |
| Unit 15 | 0.625 | 0.097 | Operational |

**Case 3**: In this case, the *N-m-u* reliability criterion is considered, where 192 contingency scenarios are defined in the system. Considering $m=5$, in each scenario five components in the outage class and one from the uncertain class are considered to be out of service. The operation cost is obtained as $1,087,022. The average load curtailment is calculated as 0.695 MWh. The obtained results advocate that considering uncertainty can offer a more resilient commitment against multiple component outages. This can be seen as the amount of load curtailment is considerably lower than considering all possible failures as an outage in same contingency scenario. This significant advantage is obtained at the expense of limited cost increase, which is less than 0.15%. The proposed approach to classifying components into three categories can therefore result in more resilient schedules.

## IV. CONCLUSION

Prediction of a component state in response to an extreme event is challenging in practice. In this paper, a multiclass SVM was proposed to categorize system components into three classes (damaged, operational, and uncertain) in response to an imminent hurricane. An artificial set of data was generated to train the SVM, as the publicly available data on the impact of hurricanes on power system components is limited. Experimental results showed the effectiveness of the proposed method in separating three classes from each other based on wind speed and the distance from the center of the hurricane. The damaged components along with the components in the uncertain class were then considered in multiple contingency scenarios of a proposed E-SCUC, which considered the simultaneous outage of multiple components under an *N-m-u* reliability criterion. Experimental results on the IEEE 118-bus test system indicated that considering uncertain components in different scenarios could potentially lead to a more resilient schedule at the expense of marginal increase in operation cost.